\documentclass[reprint, amssymb, aps, superscriptaddress, showpacs]{revtex4-1}
\usepackage{graphicx}
\usepackage{amsmath}
\usepackage{color}
\usepackage{hyperref}
\usepackage{bm}
\begin{document}

\title{Quantum Phase Transitions and Topological Proximity Effects in Graphene Nanoribbon Heterostructures}%

\author{Gufeng Zhang}%
\affiliation{International Center for Quantum Design of Functional Materials(ICQD), Hefei National Laboratory for Physical Sciences at the Microscale, University of Science and Technology of China, Hefei, Anhui, 230026, China}
\affiliation{State Key Laboratory of Surface Physics and Department of Physics, Fudan University, Shanghai, 200433, China}
\affiliation{Department of Physics, University of California, San Diego, La Jolla, CA 92093, USA}
\author{Xiaoguang Li}%
\affiliation{State Key Laboratory of Surface Physics and Department of Physics, Fudan University, Shanghai, 200433, China}
\affiliation{International Center for Quantum Design of Functional Materials(ICQD), Hefei National Laboratory for Physical Sciences at the Microscale, University of Science and Technology of China, Hefei, Anhui, 230026, China}
\affiliation{Department of Physics and Astronomy, University of Tennessee, Knoxville, TN 37996, USA}
\author{Guangfen Wu}%
\affiliation{International Center for Quantum Design of Functional Materials(ICQD), Hefei National Laboratory for Physical Sciences at the Microscale, University of Science and Technology of China, Hefei, Anhui, 230026, China}
\affiliation{Shenzhen Institutes of Advanced Technology, Chinese Academy of Sciences, Shenzhen, 518055, China}
\author{Jie Wang}%
\affiliation{International Center for Quantum Design of Functional Materials(ICQD), Hefei National Laboratory for Physical Sciences at the Microscale, University of Science and Technology of China, Hefei, Anhui, 230026, China}
\author{Dimitrie Culcer}%
\affiliation{International Center for Quantum Design of Functional Materials(ICQD), Hefei National Laboratory for Physical Sciences at the Microscale, University of Science and Technology of China, Hefei, Anhui, 230026, China}
\affiliation{School of Physics, The University of New South Wales, Sydney 2052, Australia}
\author{Efthimios Kaxiras}%
\affiliation{School of Applied Science and Engineering, Harvard University, Cambridge, MA 02138, USA}
\author{Zhenyu Zhang}%
\affiliation{International Center for Quantum Design of Functional Materials(ICQD), Hefei National Laboratory for
  Physical Sciences at the Microscale, University of Science and Technology of China, Hefei, Anhui, 230026, China}
\affiliation{School of Applied Science and Engineering, Harvard University, Cambridge, MA 02138, USA}

\begin{abstract}
  Topological insulators are bulk insulators that possess robust chiral conducting states along their interfaces with
  normal insulators. A tremendous research effort has recently been devoted to topological insulator-based
  heterostructures, in which conventional proximity effects give rise to a series of exotic physical phenomena. Here we
  establish the potential existence of \textit{topological proximity effects} at the interface between a topological
  insulator and a normal insulator, using graphene-based heterostructures as prototypical systems. Unlike conventional
  proximity effects in topological insulator based heterostructures, which refer to various phase transitions associated
  with the symmetry breaking of specific local order parameters, topological proximity effects describe the rich variety
  of quantum phase transitions associated with the global properties of the system measured by the location of the
  topological edge states. Specifically, we show that the location of the topological edge states exhibits a versatile
  tunability as a function of the interface orientation, the strength of the interface tunnel coupling between a
  topological graphene nanoribbon and a normal graphene nanoribbon, the spin-orbit coupling strength in the normal
  graphene nanoribbon, and the width of the system. For zigzag and bearded graphene nanoribbons, the topological edge
  states can be tuned to be either at the interface or outer edge of the normal ribbon. For armchair graphene
  nanoribbons, the potential location of the topological edge state can be further shifted to the edge of or within the
  normal ribbon, to the interface, or diving into the topological graphene nanoribbon. We further show that the
  topological phase diagram established for the prototypical graphene heterostructures can also explain the intriguing
  quantum phase transition reported recently in other topological-insulator heterostructures. We also discuss potential
  experimental realizations of the predicted topological proximity effects, which may pave the way for integrating the
  salient functionality of topological insulators and graphene in future device applications.
\end{abstract}


\maketitle

\section{Introduction}

The discovery of topological insulators (TIs) has revolutionized our understanding of insulating
behavior~\cite{kane05a,kane05b,BHZ,BHZe,3DTI,moore,hsieh,xia,rmphasan,rmpxlqi,nphy}. The appearance of topologically
insulating behavior is associated with a topological phase transition~\cite{kane05a,BHZ,PNAS,TPT}. For example, when the
spin-orbit coupling (SOC) exceeds a critical strength, a band inversion takes place, rendering the entire system
topologically nontrivial. Topological phase transitions do not involve symmetry breaking, but entail instead a change in
the $Z_2$ topological invariant, which may be regarded as a quantity counting the number of Dirac cones. No local order
parameters can be defined for a topological phase transition: such TI behavior, which is protected against time-reversal
invariant perturbations, is a one-particle phenomenon and is the result of SOC. The carriers at the interface between a
TI and a normal insulator are massless Dirac fermions with spin-momentum locking. At the same time, the field of TI also
provides a unique platform for studying the interplay between strong SOC and electron-electron interaction
effects~\cite{pesin,dimi}, as manifested by the existence of exotic quantum phase transitions~\cite{pesin}.

The robustness of topologically protected surface states may enable hybrid TI heterostructure systems to provide fundamental device
improvements as well as potential applications~\cite{Nchem}. Recently, by using various TI-based heterostructures, emergent
properties of topological surface states have been
demonstrated~\cite{Natural-hetero,ISGE,ferro/TI,TI/MI,TI/MI,TI/Mott,Bi/TI}. For instance, TI forms a natural
heterostructure with ordinary insulators~\cite{Natural-hetero}, which can be used to manipulate topological
states and the bulk band gap. By putting a ferromagnet on a TI, the inverse spin-galvanic effect~\cite{ISGE} and giant spin
battery effect~\cite{ferro/TI} can be realized. Interesting properties are also found in TI heterostructures with magnetic
insulators~\cite{TI/MI}, Mott insulators~\cite{TI/Mott}, and Bi(111) bilayer~\cite{Bi/TI}. In particular,
proximity effects in TI heterostructures yield novel phases of matter~\cite{proximity, stanescu, TEM, catalysis,
  hetero1, hetero2, physe, AHE}. For example, TI/superconductor heterostructures exhibit a superconducting
proximity effect offering the possibility of observing Majorana fermions~\cite{proximity} and the potential realization
of non-Abelian topological quantum computation~\cite{TQC}. The quantized electromagnetic response in a TI-ferromagnetic
material heterostructure is due to a topological magnetoelectric effect~\cite{TFT,TEM}. Other novel, technologically
important properties have also been demonstrated, such as the enhancement of the catalysis process by robust topological
surface states in Au-covered TI~\cite{catalysis}. TI-based heterostructures are thus systems of both fundamental and
practical importance~\cite{proximity, stanescu,TEM,catalysis,hetero1,hetero2}. Harnessing the robust topological surface
states entails an accurate understanding and control of their spatial location.

Recent first-principles studies of three-dimensional (3D) TI/normal insulator heterostructures have demonstrated that
the spatial location of the surface states can be shifted to the surface of the normal insulator~\cite{hetero1,hetero2},
and in a certain parameter range can even be shifted back into the TI bulk~\cite{hetero1}. Such studies suggest the
possibility of a topological phase transition induced in a normal insulator via \textit{topological proximity effects},
identified by the topological surface states leaking out into an adjacent material or moving back into the TI. Whereas
conventional proximity effects can be described using local order parameters, topological proximity effects involve a
topological phase transition for which a local order parameter cannot be defined. In light of this, a practical and
highly nontrivial issue to be addressed is the determination of the exact spatial location of the topological surface
states in such TI-based heterostructures.

\begin{figure}
\includegraphics[width=0.4\textwidth]{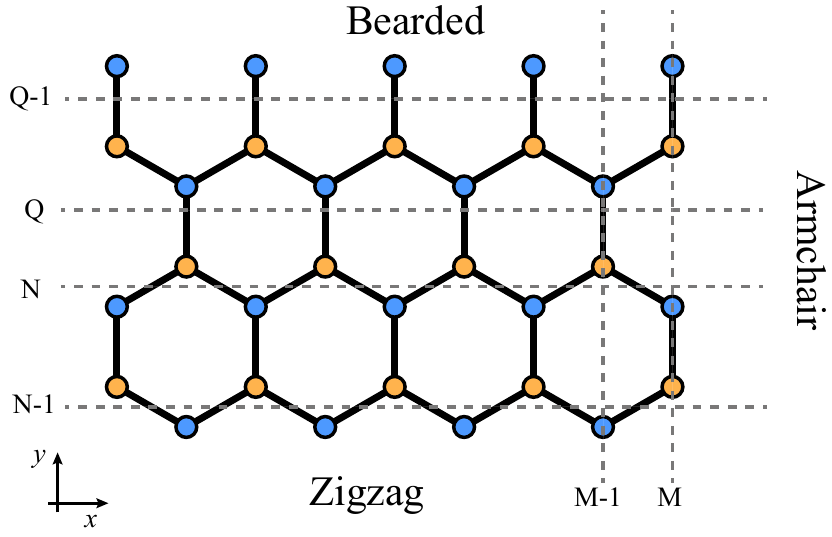}
\caption{\label{fig:scheme}(Color Online) Graphene ribbon geometry with zigzag, bearded and armchair edges. The
  dashed lines along the zigzag, armchair, and bearded edges labeled with $N$, $M$, and $Q$ indicate the $N^{th}$, $M^{th}$ and
  $Q^{th}$ unit cell in $y$, $x$ and $y$ directions, respectively. The width of a ribbon is measured in multiples of the unit cell. Ribbons
  with zigzag and bearded edged have infinite length in the $x$ direction, and for armchair edges in the $y$ direction. }
\end{figure}

In this paper, we choose graphene as a prototype to explore the general features of the topological proximity
effects. Graphene is of great interest, since it has a simple honeycomb lattice structure, while possessing numerous
intriguing topological phenomena, such as the integer quantum Hall effect~\cite{IQHE}, anomalous quantum Hall
effect~\cite{AQHE}, and fractional quantum Hall effect~\cite{FQHE}. In particular, graphene can be turned into a TI once
the intrinsic SOC reaches a large enough value~\cite{kane05a}. Given the ample energy band structure tunability of
graphene, it is convenient to use a graphene-based heterostructure as a prototype to systematically investigate the
topological proximity effects by considering several crucial parameters of the model system. We expect that the obtained
features of topological proximity effects are also suitable for other 2D or 3D TI-based systems, and may be instructive
for the design of future TI-based electronic devices.

We explore topological proximity effects on topological phase transitions in an important material class of graphene
heterostructures~\cite{GNRheter} consisting of a topological graphene nanoribbon (GNR) and a gapped normal GNR. Three
different interface orientations are investigated: zigzag, bearded, and armchair (Fig.~\ref{fig:scheme}). Unlike
conventional proximity effects, in which a phase transition is accompanied by symmetry breaking measured by a local
order parameter, the topological proximity effects introduced here surrounding topological phase transitions do not
involve a symmetry breaking process. Instead of a local order parameter, the measurement of such topological proximity
effects is the location of the topological edge states (TESs), which is determined in this study as a function of the
interface coupling strength, the SOC strength in the normal GNR, and the orientations of the interface. For three
different interface orientations: zigzag, bearded, and armchair, we demonstrate versatile tunabilities in the location
of the TES. For zigzag and bearded~\cite{bearded} GNRs, it can be tuned to be either at the interface or the outer edge
of the normal GNR. For armchair GNRs, the potential location of the TES is further enriched to be at the edge of or
within the normal GNR, at the interface, or diving into the topological GNR. Moreover, the dependence of the TES
behaviors on the interface orientation is attributed to the different locations of the Dirac points of the TES in ${\bm k}$ space for different heterostructures. Taken together, our findings illustrate the conceptual complexity as well
as richness of the topological proximity effect at TI-based heterostructures.

This paper is organized as follows: In Sec.~\ref{Method}, we introduce the tight binding Kane-Mele model for GNR
heterostructures and outline our methodology. In Sec.~\ref{result}, we show the tunability of the spatial location of
the TES as some key parameters in zigzag, bearded, and armchair GNR heterostructures. Before concluding in
Sec.~\ref{conclusion}, we mainly discuss and explain different TES behaviors for different kinds of interface
orientations in Sec.~\ref{discussion}, where potential applications and experimental realizations are also presented.

\section{Methodology}\label{Method}

We start with the Kane-Mele model~\cite{kane05a,kane05b} for GNRs. In the present work, we generalize the systems of
interest to explore the topological proximity effects in hybrid GNR heterostructures consisting of a normal and a
topological GNRs. In these systems, we focus on two central parameters: the tunnel coupling at the interface between the
GNRs, and the strength of the SOC in the normal GNR. We investigate the spatial location of the TES at the interface
between the two GNRs for three different interface orientations: zigzag, bearded, and armchair
(Fig.~\ref{fig:scheme}). These three orientations give rise to qualitatively different graphene band structures, and
consequently the proximity effects take qualitatively different forms.

For both the topological and normal GNRs, we use the same tight-binding Hamiltonian as follows, but with different
specifications on the parameters:
\begin{equation}
  H = t\sum_{\langle ij\rangle}c_i^{\dag}c_j+\sum_{i\in{a,b}} V_i c_i^{\dag}c_i+i\,\lambda_{SO}\sum_{\langle\langle ij\rangle \rangle}c^{\dag}_{i} \bm{\sigma}\cdot
  (\bm{d_{kj}}\times \bm{d_{ik}})\,c_{j}, \label{eq}
\end{equation}
where $c_i^{\dag}$($c_i$) is the electron creation (annihilation) operator on site $i$; $t$ is the nearest-neighbor
hopping; $V_{a(b)}$ is the on-site energy for the A(B) sublattice; $\lambda_{SO}$ is the intrinsic SOC connecting
next-nearest neighbors; $\bm{\sigma}$ is the Pauli matrix vector; $i$ and $j$ are two next-nearest neighbor
sites, $k$ is their unique common nearest neighbor, and the vector $\bm{d_{ik}}$ points from $k$ to $i$.

The band structure of graphene can be qualitatively changed by tuning the coupling parameters in Eq.~(\ref{eq}). When
$V_a=-V_b=V_g/2$ and $\lambda_{SO}=0$, we can obtain a trivial insulator with the band gap equals to $V_g$. On the other
hand, in a pristine graphene where we have $V_g=0$, a nontrivial insulator can be obtained with the band gap of $6
\sqrt{3}\lambda_{SO}$. Essentially, the competing between $V_g$ and $\lambda_{SO}$ determines the topological phase of
the graphene. Although the intrinsic SOC in carbon system is too weak to produce a visible nontrivial band gap
(100\,mK~\cite{SOCMac,kane05a,adatom,SOC}), many viable approaches have been proposed to enhance this. For example, the
impurity-induced, lattice driven SOC can be of at least the same order of magnitude as the atomic SOC~\cite{impurity};
certain heavy non-magnetic adatoms such as indium and thallium can enhance the intrinsic SOC gap up to room
temperature~\cite{adatom}; random adsorption of adatoms can suppress intervalley scattering, which is detrimental to the
topological phase, but does not affect the induced SOC, consequently stabilizing the topological phase in
graphene~\cite{adsorption}.

All GNR heterostructures we discuss consist of a normal and a topological GNR with the same interface orientation. The
width of the GNRs is measured in multiples of the unit cell illustrated in Fig.~\ref{fig:scheme}. In the topological
GNR, we set the SOC strength as a constant $\lambda_{SO}=0.03\,t \approx 0.08$\,eV, and the on-site energy $V_{g}=0$. In
the normal GNR, we assume a finite $V_g$ to open a trivial band gap, and a relatively small $\lambda_{SO}$, which is not
large enough to induce a topological phase transition as the GNR is isolated. The interface tunnel coupling $t_c$
appears as the nearest-neighbor hopping energy between the normal and topological GNRs; for zigzag GNR heterostructures,
the bonds with tunable coupling $t_c$ are illustrated by the red lines in Figs.~\ref{fig:Band_State}(e-h). The
heterostrctures are assumed to be infinite in the direction parallel to the interface. We thus solve an effective 1D
problem by diagonalizing the Hamiltonian in Eq.~\ref{eq} to obtain band structures and the corresponding edge states.

\section{Results}\label{result}
In this section, we report the results of band structure and spatial locations edge states in the GNR heterostructures
with different orientations (zigzag, bearded and armchair). In particular, we investigate the way the TES moves in such
systems as a result of the interplay of the interface tunneling, the spin-orbit coupling in the normal GNR and the
thickness of the normal GNR. Throughout this study, we consider the thickness of topological GNR to be a large enough
constant to avoid direct interaction between the topological edge states on the two sides of the structure. Also, we note that for simplicity, we consider a local interface coupling in this paper. Consideration of a continuously changing interface hopping strength gives qualitatively the same results.

\subsection{Zigzag graphene nanoribbon heterostructures}

\subsubsection{How interface tunneling affects the position of  topological edge states}

\begin{figure*}
\includegraphics[width=1\textwidth]{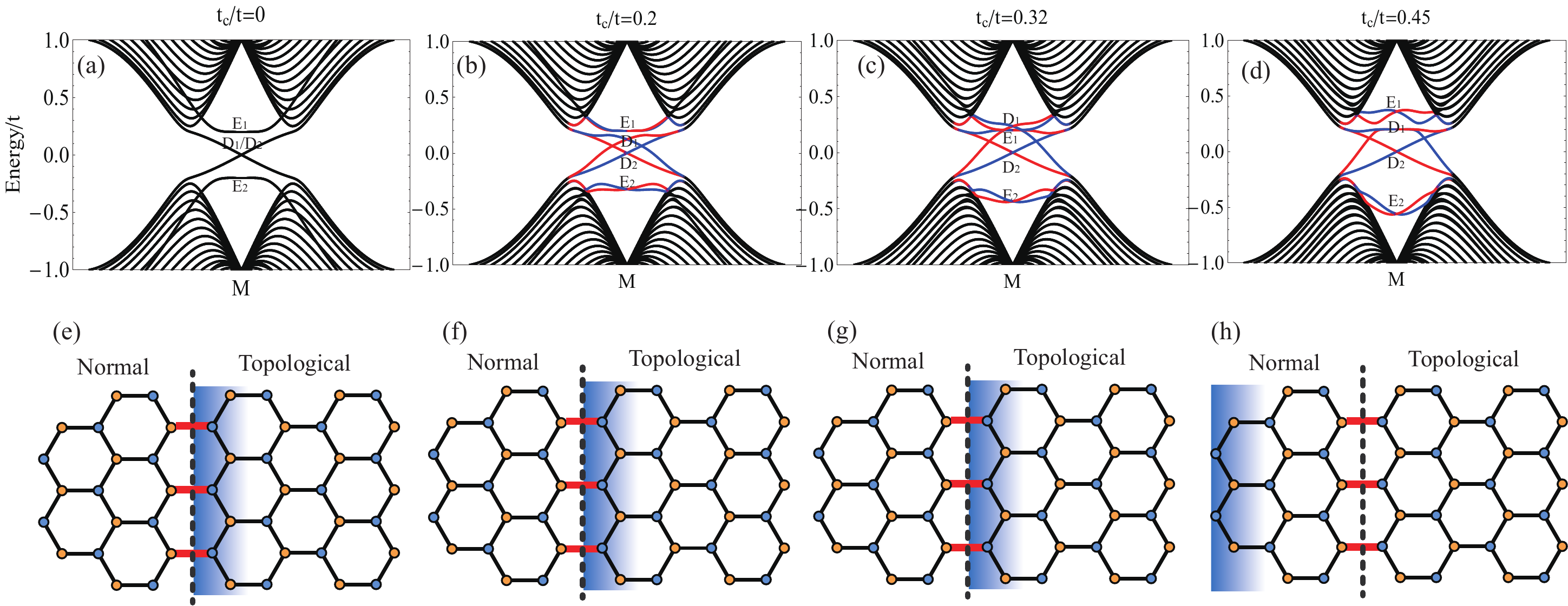}
\caption{\label{fig:Band_State}(Color Online) Band structures of zigzag edge GNR heterostructures ($W_n=3$, $W_t=30$)
  for different tunnel coupling strengths (a-d), and the corresponding spatial locations of higher energy Dirac points
  (denoted as $D_1$ in (a-d)) are shown by blue areas in (e-h). Different spin orientations are shown as spin up (blue) and
  spin down (red) near Fermi levels in band structures. The interface tunnel coupling $t_c=0$ in (a)(e), $t_c=0.2 t$ in
  (b)(f), $t_c=0.32\,t$ in (c)(g), $t_c=0.45\,t$ in (d)(h). For the normal GNR, $V_g=0.1\,t$ and $\lambda_{SO}=0$; for the
  topological GNR, $V_{g}=0$ and $\lambda_{SO}=0.03\,t$. $E_1$ and $E_2$: the bands of the trivial edge states
  originated from the normal GNR. $D_2$: the other Dirac point whose spatial location is far away from the
  interface. Dashed lines in (e-h) indicate the interface of the heterostructures, and the red bonds show the interface
  tunnel coupling ($t_c$).}
\end{figure*}

We first investigate GNR heterostructures with zigzag edges. The heterostructure consists of a normal GNR with width
$W_n=3$, and a topological GNR with width $W_t=30$. Originally, the normal zigzag GNR possesses gapless edge
states~\cite{GNR}. To distinguish between the topological and trivial edge states, we open a band gap in the normal GNR
by adding the different on-site energy for the two sublattices. In Figs.~\ref{fig:Band_State}(a-d), we exhibit a series
of band structures of systems with the different interface coupling $t_c$, while the fixed on-site energy $V_a=
-V_b=0.1\,t$ and SOC strength $\lambda_{SO}=0$ in the normal GNR. $E_1$ and $E_2$ represent the energy of trivial edge
states, which locate separately at the edges of the normal GNR, and the \textit{Dirac points} (denoted as $D_1$ and
$D_2$) describing the band crossing of the TESs to emphasize the linear dispersion around the corresponding $k$
point. We also show spin orientations near the Fermi level in Figs.~\ref{fig:Band_State}(b-d) with spin up in blue and
spin down in red. Based on our definition, the electrons of $E_2$ and $D_1$ are located near the interface between the
normal and topological GNRs, and therefore are more sensitive to the tunnel coupling $t_c$. As we gradually increase
$t_c$ as illustrated in Fig.~\ref{fig:Band_State} from left to right, both the energies and spatial locations of these
edge states will change accordingly. To see the evolution of the states clearly, we analyze the transition as follows:

\begin{itemize}
\item $t_c/t=0$ (Figs.~\ref{fig:Band_State}(a)(e)). The two GNRs are detached, and the two Dirac points are
  degenerate. The state $D_1$ is located at the interface of the heterostructure, as shown by the blue
  area in Fig.~\ref{fig:Band_State}(e).
\item $t_c/t=0.2$ (Figs.~\ref{fig:Band_State}(b)(f)). As $t_c$ increases, the interaction between bands $D_1$ and $E_2$
  results in the energy of $D_1$ increasing and that of $E_2$ decreasing. But the spatial location of $D_1$ does not
  change.
\item $t_c/t=0.32$ (Figs.~\ref{fig:Band_State}(c)(g)). The energy of $D_1$ increases to slightly exceed $E_1$. The
  spatial location of $D_1$ stays at the interface.
\item $t_c/t=0.45$ (Figs.~\ref{fig:Band_State}(d)(h)). The topological phase transition happens when the bands $E_1$ and
  $D_1$ detach. Now $D_1$ becomes the \textit{flat} band with energy lower than $E_1$, and its spatial location moves to
  the outer edge of the normal GNR (Fig.~\ref{fig:Band_State}(h)).
\end{itemize}

After the phase transition, further increasing $t_c$ will not change the energy and spatial location of $D_1$, but will
keep enlarging the energy of $E_1$ and reducing the energy of $E_2$ toward bulk states, implying that these two states
are spatially more close now. In the whole process, the energy of $D_2$ does not change, because its spatial location is
far away from the interface. For very large $t_c$, $E_1$ and $E_2$ are mixed with bulk states, and we can see only two
TESs located separately at two edges of the whole heterostructures. The shift of TESs from the interface to the outer
edge of the normal GNR indicates that the heterostructure in its entirety becomes an expanded 2D TI via the topological
proximity effect.

\begin{figure}
\includegraphics[width=0.45\textwidth]{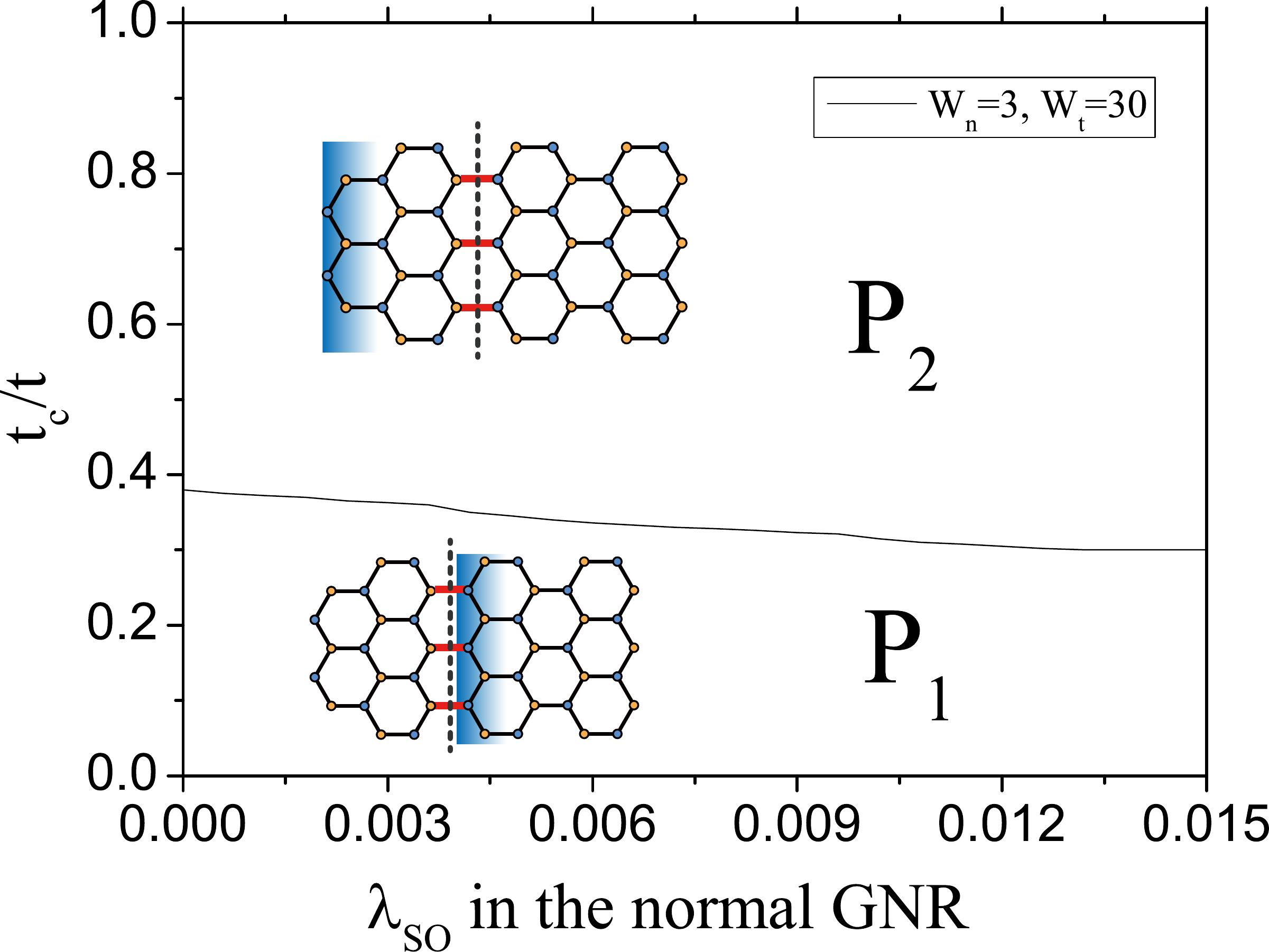}
\caption{\label{fig:phase1}(Color Online) Phase diagram for zigzag GNR heterostructures, spanned by the tunnel
  coupling ($t_c$) and SOC in the normal GNR. The black solid line indicates the boundary of $P_1$ and $P_2$ phases with
  $W_n=3$, $W_t=30$. Insets are the illustrations of the spatial locations of the TESs. $P_1$ phase: the TES is located
  at the interface. $P_2$ phase: the TES is located at the outer edge of the normal GNR.}
\end{figure}

\subsubsection{How spin-orbit coupling affects the position of topological edges states}

Qualitatively similar pictures are observed when we include a finite SOC in the normal GNR. Under the influence of both
the tunnel coupling and SOC, the system still has two phases. As shown in the phase diagram Fig.~\ref{fig:phase1}, the
phase transition occurs at the boundary of the $P_1$ and $P_2$ areas. Specifically, for given values of $t$ and
$\lambda_{SO}$, $t_c$ separating the two phases decreases monotonically as the SOC in the normal GNR increases. In other
words, the SOC in the normal GNR helps the transition happen. We can understand this physical picture by considering the
limiting case when we apply a large enough SOC in the normal GNR to induce the topological phase transition without the
presence of other topological proximities. In this case, the TES will not appear at the interface ($P_1$ phase) for any
finite $t_c$, but at the edges of the whole heterostructure ($P_2$ phase), because both GNRs are topologically
nontrivial.

\subsubsection{Whole phase diagram of the position of  topological edge states}

\begin{figure}
\includegraphics[width=0.5\textwidth]{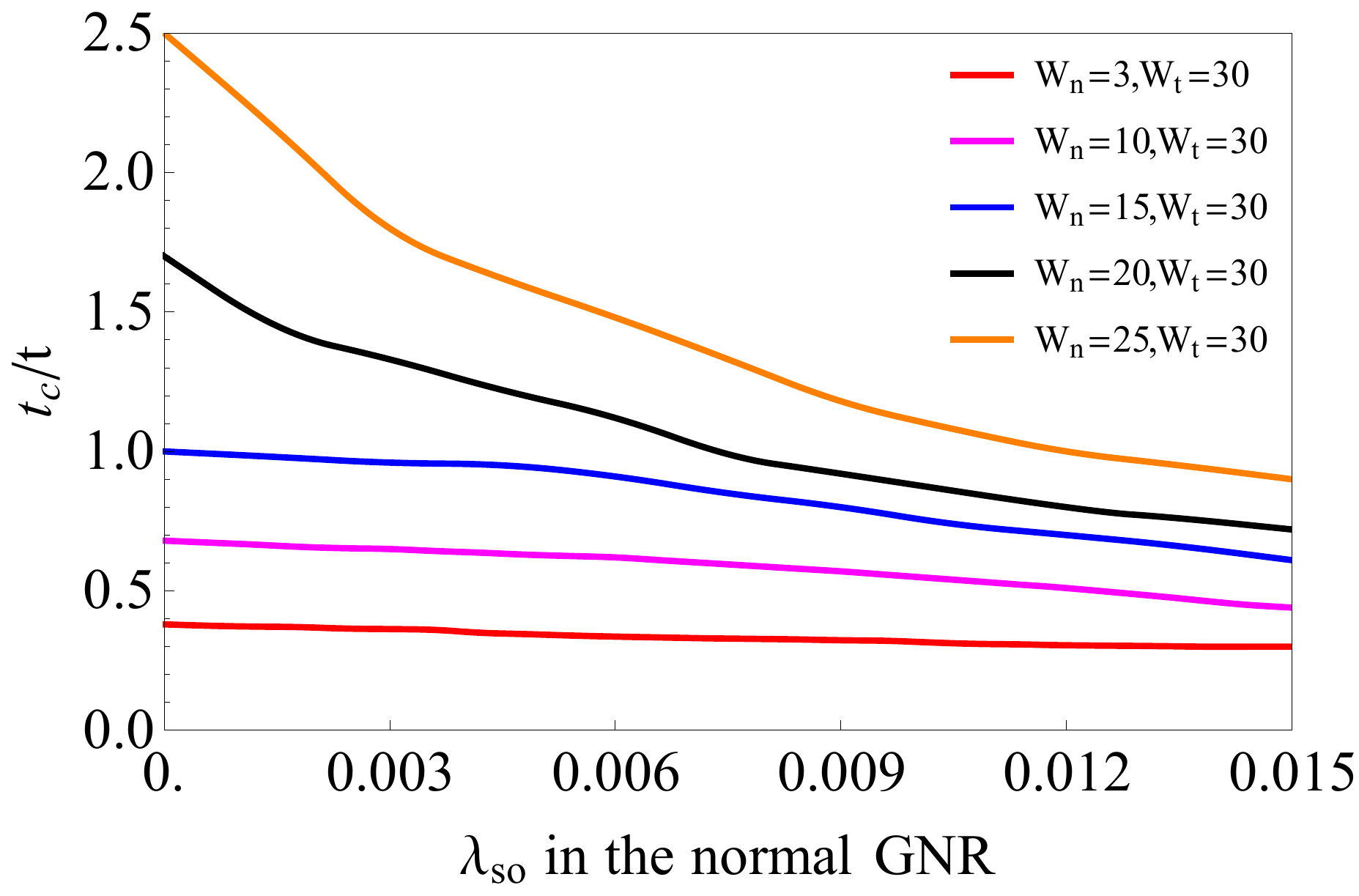}
\caption{\label{fig:thickz}(Color Online) Phase diagram for zigzag GNR heterostructures with different normal GNR
  widths, spanned by the tunnel coupling ($t_c$) and SOC in the normal GNR. The lines indicate the boundary of the $P_1$ and
  $P_2$ phases, which are defined the same as in Fig.~\ref{fig:phase1}. We use different colors for various widths, as
  shown in the inset.}
\end{figure}

Furthermore, we discuss the way the width of the normal GNR affects the phase transition. Fig.~\ref{fig:thickz} shows
the phase diagram for normal GNR with different widths $W_n=3, 10, 15, 20, 25$. We see that for the same
SOC in the normal GNR, the wider one requires a larger tunnel coupling $t_c$ to induce the transition from $P_1$ to
$P_2$. This is because the phase transition essentially needs the coupling between the states at the interface and outer
edge of the normal GNR, and this coupling becomes weaker as the width increases. So a larger tunnel coupling $t_c$ is
required to \textit{propagate} the TESs to further position. In an limiting case where the normal ribbon has infinite
width, topological proximity effect will not occur for any finite tunnel coupling $t_c$.

In summary for the zigzag GNR heterostructures, two phases exist with different tunnel coupling and SOC in the normal
GNR: the TES may be located at the interface or at the outer edge of the normal GNR.

\subsection{Bearded graphene nanoribbon heterostructures}

\begin{figure}
\includegraphics[width=0.4\textwidth]{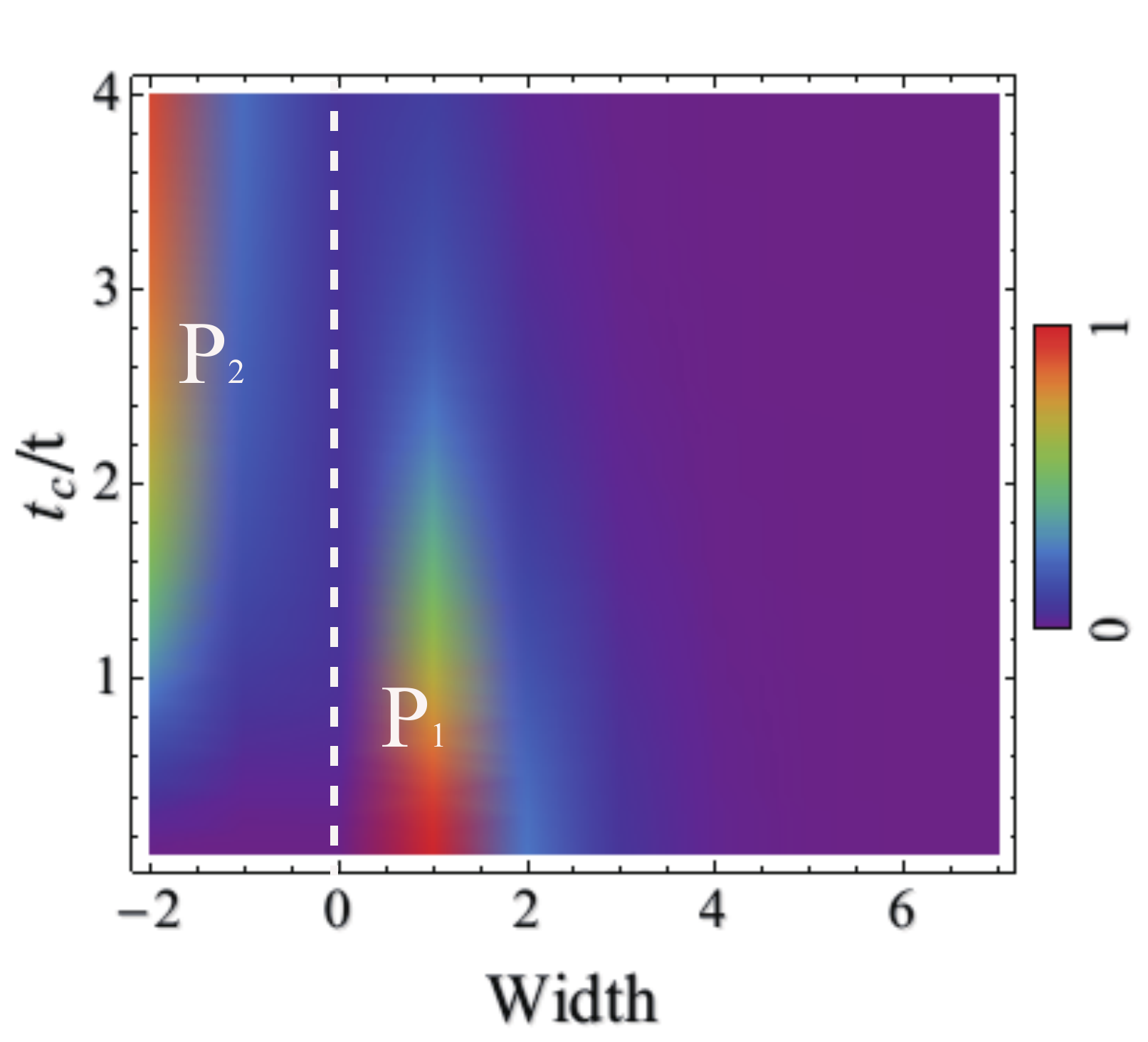}
\caption{\label{fig:bearded}(Color online) Phase diagram of the TES location without SOC in the normal GNR. The
  heterostuctures consist of a normal GNR with width $W_n=3$ and a topological GNR with width $W_t=30$. The dashed line
  indicates the interface of the heterostructures. $P_1$ phase: the TES is located at the interface. $P_2$ phase: the TES
  is located at the outer edge of the normal GNR.}
\end{figure}

We now consider bearded edge graphene nanoribbon~\cite{bearded} heterostructures. Similar to zigzag GNR, a bearded GNR
has gapless trivial edge states. After turning on the SOC, a Dirac point emerges at $\Gamma$ point
(k=0)[Fig.~\ref{fig:18}(b)]. We show the topological phase diagram of the bearded edge GNR heterostructure in
Fig.~\ref{fig:bearded}. The system consists of a normal GNR with width $W_n=3$ and topological GNR $W_t=30$. As one
increases the interface tunnel strength ($t_c$) between the heterostructures, the TES moves from the interface ($P_1$
phase) to the outer edge of the normal ribbon ($P_2$ phase)[Fig.~\ref{fig:bearded}]. Furthermore, switching on the SOC
in the normal GNR still preserves the two phases for the TES. The influence of SOC in normal GNR on the topological
proximity effect is the same as in the zigzag case.

\subsection{Armchair graphene nanoribbon heterostructures}

\begin{figure*}
\includegraphics[width=1\textwidth]{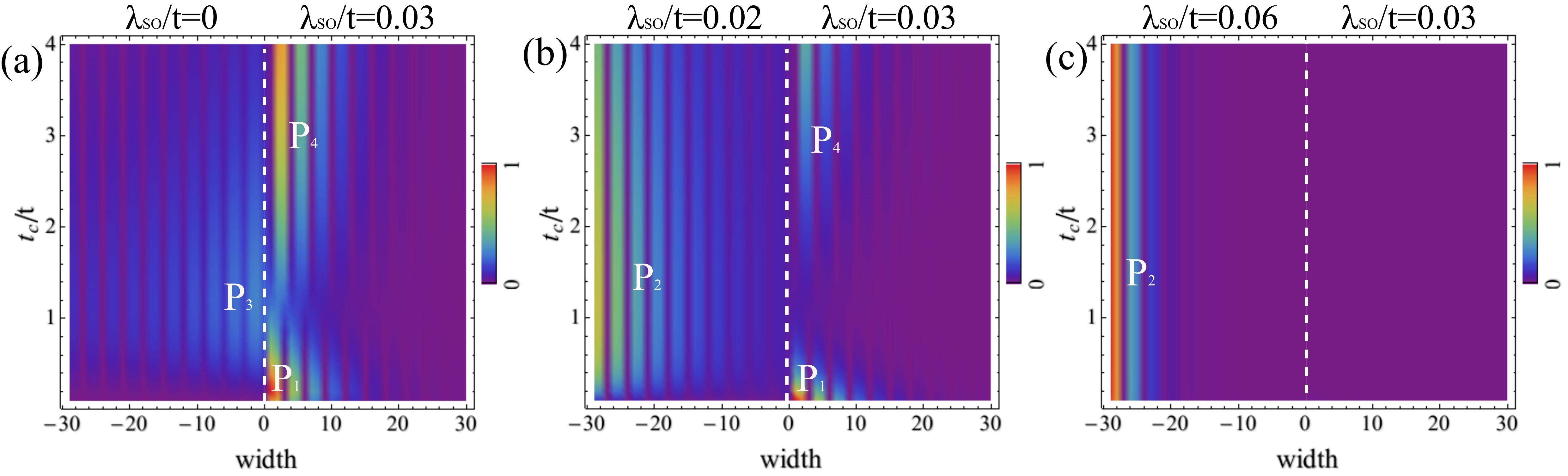}
\caption{\label{fig:phase2}(Color online) Phase diagram of the TES location for armchair GNR heterostructures with a
  normal GNR width $W_n=30$ on the left and a topological GNR width $W_t=60$ on the right. The SOC strength
  $\lambda_{SO}=0.03\,t$ in the topological GNR and is 0 (a), $0.02\,t$ (b) and $0.06\,t$ (c) in the normal GNR. The
  on-site energy in the normal GNR is $V_g=0.03\,t$. The dashed lines indicate the interface location. Four phases
  appear for the TES as the tunnel coupling ($t_c$) increases: $P_1$ phase: the TES is located at the interface; $P_2$
  phase: the TES at the outer edge of the normal GNR; $P_3$ phase: the TES in the bulk of the normal GNR; $P_4$ phase:
  the TES is located one unit cell into the topological GNR.}
\end{figure*}

Armchair edge GNR heterostructures differ a lot from both zigzag and bearded edge systems in that a pristine
(i.e.~without SOC) armchair GNR does not have trivial edge states~\cite{GNR}. We first investigate the spatial location
of the TES as the tunnel coupling $t_c$ increases for the fixed SOC in the normal GNR. Fig.~\ref{fig:phase2} displays
our calculation results for three SOC values in the normal GNR: (a) no SOC; (b) moderate SOC ($\lambda_{SO}/t=0.02$) and
(c) strong SOC ($\lambda_{SO}/t=0.06$), which is large enough to induce topological phase transition in the normal GNR
at $t_c=0$. In contrast with the zigzag and bearded cases, the armchair GNR heterostructure shows two additional phases
in the strong tunnel coupling regime, $P_3$, where the density of the TES peaks in the normal GNR, and $P_4$, where the
TES is re-located one unit cell back inside the topological GNR. With no SOC in the normal GNR (a), the location of the
TES shifts from $P_1$ through $P_3$ to $P_4$ as the coupling increases, while for a moderate SOC (b), the TES can move
to the outer edge of the normal GNR through $P_2$ instead of $P_3$ during the evolution. In the limiting case of the
strong SOC, the original normal GNR becomes topologically nontrivial. Thus the whole system becomes a 2D TI. With any
finite $t_c$, the TES will only exist at the boundary of the whole system as shown in Fig.~\ref{fig:phase2}. We note
that neither the $P_3$ nor the $P_4$ phase exists for the zigzag or bearded GNR, and the appearance of the $P_4$ phase
is consistent with \textit{ab initio} work reported recently~\cite{hetero1}.

\begin{figure*}
\includegraphics[width=1\textwidth]{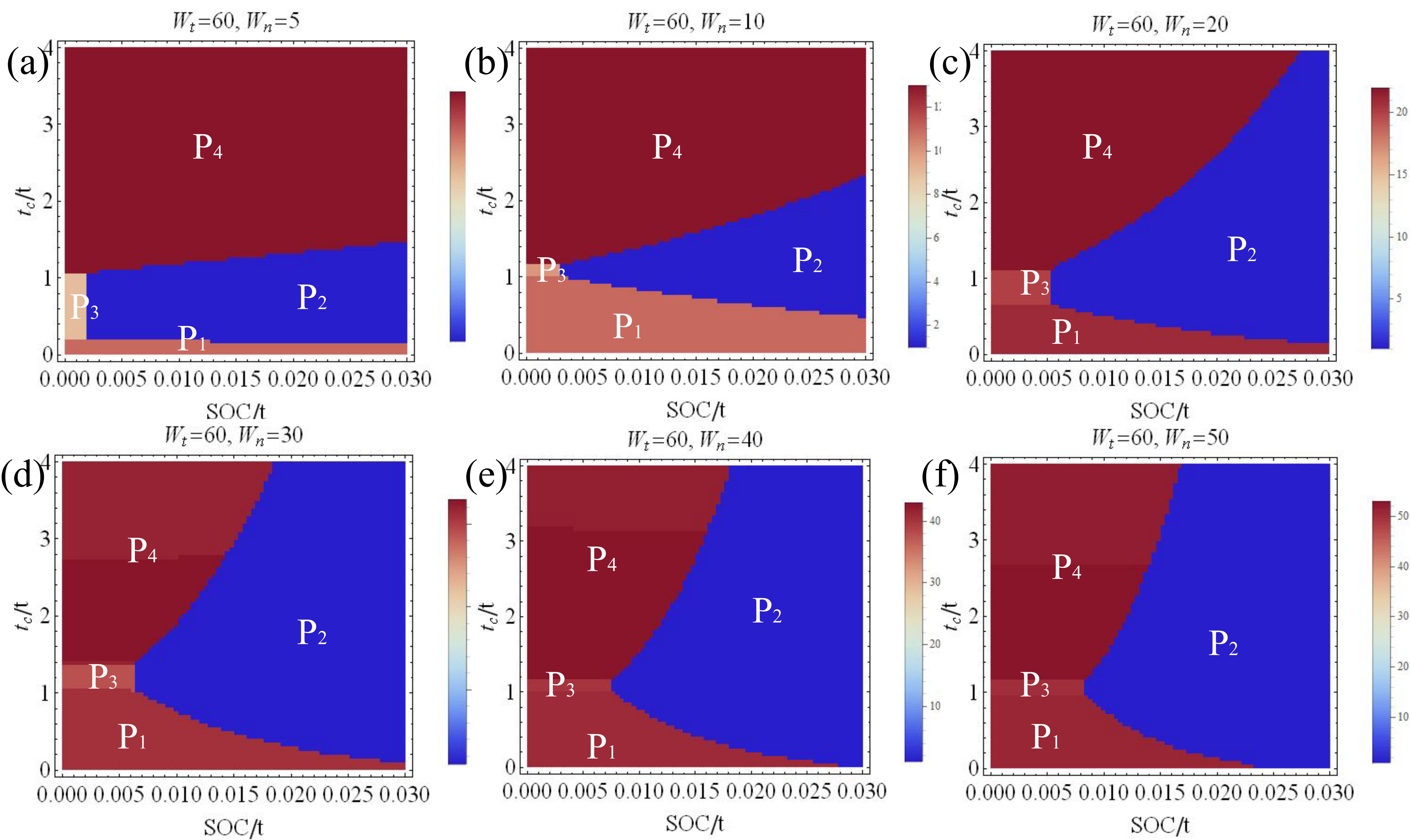}
\caption{\label{fig:armthi}(Color online) Phase diagram for the armchair GNR heterostructures with a topological GNR
   width $W_t=60$ and a normal GNR  width $W_n=5$ (a), $W_n=10$ (b), $W_n=20$ (c), $W_n=30$ (d), $W_n=40$ (e),
  $W_n=50$ (f). All diagrams are spanned by the tunnel coupling ($t_c/t$) and SOC in the normal GNR. The four phases are
  defined same as in Fig.~\ref{fig:phase2}.}
\end{figure*}

Next, we investigate a series of phase diagram by considering the widths of the normal GNR. Fig.~\ref{fig:armthi}
displays the phase diagrams as a function of the tunnel coupling and SOC in the normal GNR for the heterostructures with
various normal GNR width $W_n$ and fixed topological GNR width $W_t=60$. Generally speaking, as $t_c$ increases, we see
three possible evolution routines for a fixed SOC in the normal GNR: $P_1 \rightarrow P_2 \rightarrow P_4$, $P_1
\rightarrow P_3 \rightarrow P_4$, and $P_2$, which correspond to the three subfigures in Fig.~\ref{fig:phase2}. From
Fig.~\ref{fig:armthi}, we see clearly that the phase space of $P_2$ extends as the width increases. For a wider normal
GNR and large enough SOC, the TESs will not move one unit cell into the topological GNR ($P_4$ phase), but
continue to reside at the outer edge of the normal GNR ($P_2$ phase). In such a case, the whole heterostructure acts as a
topological insulator which has the TESs on both sides.

\section{Discussion}\label{discussion}

\begin{figure*}
\includegraphics[width=1\textwidth]{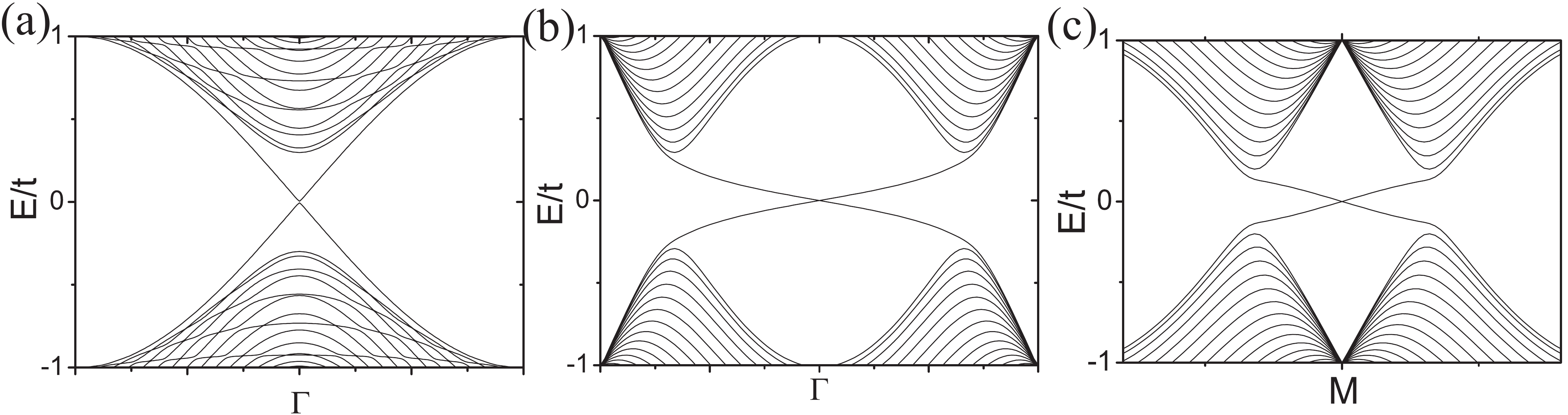}
\caption{\label{fig:18}(Color online) Band structures of an armchair (a), bearded (b), and zigzag (c) topological GNRs
  with the widths  $W_t=30$.}
\end{figure*}

The movement of the TES to different locations is a manifestation of the complexity of the topological proximity
effect. As the interface tunnel coupling becomes stronger, it gets easier for the SOC on the topological side of the junction
to leak into the part without SOC, which gives rise to an effective SOC in the normal GNR. However, the phase diagram of
the TES locations is different for of the GNR heterostructures with different orientations. The existence of the trivial
edge states in zigzag and bearded heterostructures plays an important role in causing this qualitative
difference. Fig.~\ref{fig:18} displays the band structures of the isolated topological GNRs with different edge
orientations. We note that the Dirac points for the armchair, bearded and zigzag edge GNRs appear at different $k$
points. For the zigzag (bearded) edge GNR, the Dirac point at M($\Gamma$) point is energetically far away from the bulk
states. Consequently, the TES at the M($\Gamma$) points prefers to mix with the edge states in normal GNR, which is
energetically close to the TES, therefore this rules out the existence of the $P_3$ and $P_4$ phases. However, for the
armchair edge GNR, the energy of the Dirac point is close to the bulk bands and the trivial edge state does not
exist. As a consequence, it is possible for the TES to interact with the bulk states and move into the bulk, providing
the tow more phases $P_3$ and $P_4$ related to the bulk states.

In reality, most known TIs have similar band structures to armchair GNRs, namely, the Dirac points of the TES are
located at the same $k$ point as the bulk band gap (Fig.~\ref{fig:18}(a)). We thus expect that the phase diagram of the
armchair GNR heterostructure is also applicable to the similar TI-based heterostructures. In a recent \textit{ab initio}
study, Wu $et$ $al$~\cite{hetero1} have observed in 3D TI/normal insulator heterostructures that the spatial location of
the topological surface states can be located at the interface, shifted to the surface of the normal insulator, or back
into the TI bulk, which corresponds to $P_1$, $P_2$, and $P_4$ in our phase diagram, respectively. On the other hand,
the new discovered $P_3$ phase in this study can be expected in the TI/normal insulator heterostructures with a
relatively thick normal insulator layer. For the zigzag GNR heterostructures, we also expect an analogy between the
prototypical graphene system and other 3D TI/normal insulator heterostructures with the similar band structures.

At this point, it is worthwhile to emphasize that conventional proximity effects involve the order parameter of a
broken-symmetry phase of a host material leaking into an adjacent material, which is driven into a broken symmetry state
of the host material as well. In the topological proximity effects demonstrated here, modulations of the SOC and
interface tunnel coupling shift the boundary between a normal insulator and a TI, accompanied by topological phase
transitions. A fascinating feature of the topological proximity effects is the dual-proximity nature: the location of
the TES can be switched back and forth between the two materials, including placing the topological phase boundary
inside an otherwise structurally homogenous material such as the normal insulator or the TI. One can regard a conventional
proximity effect as a spatial extension of a broken symmetry state. In contrast, the topological proximity effect refers
to inducing chiral surface states in an adjacent material as well as controlling their location, again, without symmetry
breaking.

Aside from the conceptual advances, the present study may also offer new opportunities in developing spintronic devices
and quantum computing. For example, the systems proposed in this work can be used to induce chiral spin-polarized states
in a 2D graphene slab at will, which constitutes an ON/OFF switch based on the TES, and can be regarded as a qubit. Such
a switch can be controlled by the SOC or the tunnel coupling. As another example, we can further regard the TES as a
current loop, and join a square slab of a topological GNR with a square slab of a normal GNR. Whereas the topological
side is always in the ON state, the state of the normal insulator side can be modulated. A spin polarization may then be
induced in this qubit by coupling it to a ferromagnet, which would enable control of the quantum anomalous Hall effect,
and would also serve as a spin injector into the normal GNR.

Although the predicted topological proximity effects are not limited for the graphene systems, as a very promising
material for future electronic devices, we would like to briefly discuss the potential experimental realization of the
topological proximity effects predicted in graphene. Firstly, as we mentioned before, even though the intrinsic SOC in
pristine graphene is commonly known to be quite weak, many approaches have been proposed to enhance it
\cite{adatom,impurity,adsorption}. Recent experiment progresses~\cite{Colossal} reported colossal enhancement of SOC to 2.5meV in weakly hydrogenated graphene. Secondly, the coupling between a normal and a topological GNR can be effectively tuned by substrate steps, as demonstrated recently in Ref.~\cite{step}, where the resistance from the steps on the SiC
substrate was found to rise due to the abrupt variation in potential and doping as the graphene extends over a
step. Lastly, the recently predicted 2D organic topological insulators\cite{OTI1,OTI2} with a hexagonal lattice and much larger SOC would offer another platform to implement the topological proximity effect predicted here. These and other alternative candidate structural systems may therefore provide test grounds for physical realization of topological phase transitions.

\section{Conclusions}\label{conclusion}

We have demonstrated the existence of topological proximity effects in GNR heterostructures consisting of a normal and a
topological GNR under a variety of experimentally relevant circumstances. For different types of edges - zigzag,
bearded, and armchair - the location of the TES is a function of the interface tunnel coupling, the SOC strength in the
normal GNR and also the width of the normal GNR, demonstrating a rich quantum phase diagram. These findings pave the way
for designing next-generation quantum devices that integrate the functionality of graphene and TI. We also stress that
the novel topological proximity effects demonstrated here using the prototypical systems of GNR heterostructures are
conceptually also applicable to heterostructures consisting of a normal insulator and a 3D
TI~\cite{Natural-hetero,hetero1,hetero2}.
\\

\begin{acknowledgments}
  This work was supported by NSF of China (Grant Nos. 91021019 and 11034006), National Basic Research Program of China
  (Grant No. 2011CB921801), USDOE (Grant No. DE-FG03-02ER45958), and USNSF (Grant No. 0906025).
\end{acknowledgments}

\bibliographystyle{unsrt}

\end{document}